\documentclass{ws-procs975x65}
\usepackage[pdftex]{hyperref}

\def\bPhi{\boldsymbol{\Phi}}
\def\boldf{\boldsymbol{f}}
\def\be{\begin{equation}}
\def\ee{\end{equation}}
\def\beq{\begin{eqnarray}}
\def\eeq{\end{eqnarray}}

\begin{document}

\title{The formalism of invariants in scalar-tensor and \\ multiscalar-tensor theories of gravitation}
\author{Laur J\"arv$^*$, Piret Kuusk, Margus Saal, Ott Vilson}

\address{Institute of Physics, University of Tartu,\\
Tartu, 50411, Estonia\\
$^*$E-mail: laur.jarv@ut.ee
}

\begin{abstract}
We give a brief summary of the formalism of invariants in general scalar-tensor and multiscalar-tensor gravities without derivative couplings. By rescaling of the metric and reparametrization of the scalar fields, the theory can be presented in different conformal frames and parametrizations. Due to this freedom in transformations, the scalar fields themselves do not carry independent physical meaning (in a generic parametrization). However, there are functions of the scalar fields and their derivatives which remain invariant under the transformations, providing a set of physical variables for the theory. We indicate how to construct such invariants and show how the observables like parametrized post-Newtonian parameters and characteristics of Friedmann-Lema\^itre-Robertson-Walker cosmology can be neatly expressed in terms of the invariants.

\end{abstract}

\keywords{scalar-tensor gravity; multiscalar-tensor gravity; parameterized post-Newtonian parameters; Friedmann cosmology; general relativity limit.}

\bodymatter

\section{Introduction}

Scalar-tensor gravity (STG) introduces a scalar field  that is nonminimally coupled to curvature and
thus can be interpreted as an additional mediator of gravitational
interaction besides the usual metric tensor.
It is well known that by rescaling of the metric and reparametrization of the scalar field, the theory can be presented in different conformal frames and parametrizations.\citep{flanagan} 
Despite
an extensive use of this property as a convenient calculational tool, there lingers a conceptual issue of what is the precise relation of different frames and parametrizations to the observable world and to each other.
One may interpret the change of a conformal frame and reparametrization as a change of coordinates in some abstract generalized field space. 
From this point of view the confusion arises from the fact that the theory has not been formulated in a covariant way with respect to that abstract space. 

One possible path to proceed has been to introduce conformally invariant variables.\cite{catena, postma,wetterich2015}
In this review we summarize our work on quantities in STG\cite{meie2014talv, meie2014, meie2015, ott2015} and multiscalar-tensor gravity (MSTG)\cite{mstg2015} that are invariant not only under the conformal rescaling but also under the scalar field redefinition. Three basic invariants and invariant metric enable us to formulate the theory and express physical observables independently of the choice of frame and parametrization.

\section{Parametrizations of Scalar-Tensor Gravity}

The most generic ``first generation'' STG where the scalar field $\Phi$ has no derivative couplings, is described by four arbitrary functions ${\mathcal A}(\Phi)$, ${\mathcal B}(\Phi)$, ${\mathcal V}(\Phi)$, $e^{2\alpha(\Phi)}$
in the action (where $\chi$ stands for the matter fields)\cite{flanagan, meie2014} 
\be
\label{fl_moju}
S = \frac{1}{2\kappa^2}\int d^4x\sqrt{-g}\left\lbrace {\mathcal A}(\Phi)R-
{\mathcal B}(\Phi)g^{\mu\nu}\nabla_\mu\Phi \nabla_\nu\Phi - 2\ell^{-2}{\mathcal V}(\Phi)\right\rbrace 
 + S_{\mathrm m}\left[e^{2\alpha(\Phi)}g_{\mu\nu},\chi\right] \,.
\ee
The two dimensionful constants $\kappa^2$, $\ell$ have been chosen such as to make $\Phi$ dimensionless.
Let us assume that
$0 < {\mathcal{A}}  < \infty$, $0< 2 {\mathcal{A}} {\mathcal{B}} + 3 \left({\mathcal{A}}^{\prime}\right)^2$, 
$0 \leq {\mathcal V} < \infty$, 
$|{\alpha}|  < \infty$.

By conformal rescaling and scalar field redefinition one can fix two out of the four functions to get different frames and parametrizations. For instance 
Jordan frame Brans-Dicke-Bergmann-Wagoner parametrization (JF BDBW) is obtained by
${\mathcal A} = \Psi$, ${\mathcal B} = \frac{\omega(\Psi)}{\Psi}$, ${\mathcal V}= {\mathcal V}(\Psi)$, $\alpha = 0$. 
On the other hand Einstein frame canonical parametrization (EF canonical) 
has
${\mathcal A} = 1$, ${\mathcal B} = 2$, 
${\mathcal V}={\mathcal V}(\varphi)$, $\alpha = \alpha(\varphi)$.

\section{Transformation Rules and Invariants}

Under conformal rescaling of the metric and scalar field reparametrization
\be
g_{\mu\nu}= e^{2\bar{\gamma}(\bar{\Phi})}\bar{g}_{\mu\nu}, \qquad \qquad \Phi = \bar{f}(\bar{\Phi}) \,,
\label{transformations}
\ee	
the functions transform as\cite{flanagan}
\beq
	\bar{\mathcal{A}}(\bar{\Phi}) &=& e^{2\bar{\gamma}(\bar{\Phi})}
	{\mathcal A} \left( {\bar f}( {\bar \Phi})\right) \,,
	\nonumber \\
	{\bar {\mathcal B}}({\bar \Phi}) &=& e^{2{\bar \gamma}({\bar \Phi})}\left( 
	\left(\bar{f}^\prime\right)^2{\mathcal B}\left(\bar{f}(\bar{\Phi})\right) -
	 6\left(\bar{\gamma}^\prime\right)^2{\mathcal A}\left(\bar{f}(\bar{\Phi})\right) -
	  6\bar{\gamma}^\prime\bar{f}^\prime \mathcal{A}^\prime \right) \,,
	  \nonumber \\
	\bar{{\mathcal V}}(\bar{\Phi}) &=& e^{4\bar{\gamma}(\bar{\Phi})} \, {\mathcal V}\left(\bar{f}(\bar{\Phi})\right) \,, \nonumber \\
	\bar{\alpha}(\bar{\Phi}) &=& \alpha\left(\bar{f}(\bar{\Phi})\right) + \bar{\gamma}(\bar{\Phi})\,.
	\label{fl_fnide_teisendused}
\eeq

We can inspect these rules to find combinations which remain invariant. It is convenient to write out three basic independent quantitites, invariant under rescaling and reparametrization:\cite{meie2014}
\beq
\label{I_1}
\mathcal{I}_1(\Phi) &\equiv& \frac{e^{2\alpha(\Phi)}}{\mathcal{A}(\Phi)} \,, \qquad \qquad
\label{I_2}
\mathcal{I}_2(\Phi) \equiv \frac{\mathcal{V}(\Phi)}{\left(\mathcal{A}(\Phi)\right)^2} \,, \\
\label{I_3}
\mathcal{I}_3(\Phi) &\equiv& \pm \int \left(\frac{2 {\mathcal{A}} {\mathcal{B}} + 3 \left({\mathcal{A}}^{\prime}\right)^2}{4{\mathcal{A}}^2} \right)^{\frac{1}{2}} d\Phi \,.
\eeq
Here ${\mathcal I}_1(\Phi) \not\equiv const$ signals nonminimal coupling, i.e.\ the constant ${\mathcal I}_1$ case is equivalent to a scalar field minimally coupled to curvature. Next, ${\mathcal I}_2(\Phi) \not\equiv 0$ means nonvanishing potential, related to the fact that a scalar without mass and self-interactions remains so in all frames and parametrizations. Finally, 
$\left( {\mathcal I}_3'(\Phi) \right)^2 = \frac{2 \omega(\Psi) +3}{4 \Psi^2}$ is a useful quantity that frequently appears in various equations and formulas.

On the basis of the three basic invariants it is possible to define infinitely many more invariants by (a) taking some function of them, 
$\mathcal{I}_i \equiv \mathfrak{f}( \mathcal{I}_j )$, 
(b) forming a quotient of the derivatives with respect to the scalar field, 
$\mathcal{I}_m\equiv\frac{\mathcal{I}_k^\prime}{\mathcal{I}_l^\prime} \equiv \frac{d I_k}{d I_l}$, 
(c) taking an integral 
$\mathcal{I}_r \equiv \int \mathcal{I}_n\mathcal{I}_p' d\Phi$. 
So, for example\cite{meie2014}
\be
 \mathcal{I}_4 \equiv  \frac{\mathcal{I}_2}{\mathcal{I}_1^2} = \frac{\mathcal{V}}{e^{4\alpha}}\,, \qquad
 \label{I_5}
 \mathcal{I}_5 \equiv \left(\frac{\mathcal{I}_1^\prime}{2\mathcal{I}_1\mathcal{I}_3^\prime}\right)^2 = \frac{\left(2\alpha^\prime\mathcal{A} - \mathcal{A}^\prime\right)^2}{2\mathcal{A}\mathcal{B} + 3\left(\mathcal{A}^\prime\right)^2} \,.
\ee
are also invariants, i.e.\ do not change under the transformations (\ref{transformations}).

\section{Action in Terms of Invariants}

One may also introduce an additional invariant object involving the metric. For instance taking
$\hat{g}_{\mu\nu} \equiv \mathcal{A}(\Phi)g_{\mu\nu}$ and using it to define invariant curvature,  
the action (\ref{fl_moju}) can be expressed via invariant quantities as
\be
S = \frac{1}{2\kappa^2}\int d^4x\sqrt{-\hat{g}}\left\lbrace \hat{R} - 2\hat{g}^{\mu\nu}\hat{\nabla}_\mu\mathcal{I}_3 \hat{\nabla}_\nu\mathcal{I}_3 - 2\ell^{-2}\mathcal{I}_2 \right\rbrace 
\label{invariant_action_functional} 
+ S_m\left[ \mathcal{I}_1 \hat{g}_{\mu\nu} , \chi \right] \,.
\ee
Taking $\mathcal{I}_1(\mathcal{I}_3)$ and $\mathcal{I}_2(\mathcal{I}_3)$, and varying w.r.t.\ $\mathcal{I}_3$ and $\hat{g}_{\mu\nu}$ gives invariant field equations that are equivalent to the known equations in particular frames and parametrizations.\cite{meie2014}

\section{PPN Parameters in Terms of Invariants}

Physically observable quantities should be  independent of the choice of frame and parametrization, hence they should be expressible as functions of the invariants.
Indeed, the effective Newton's constant and parameterized post-Newtonian (PPN) parameters,
computed in case of the point mass source for JF BDBW parametrization for general kinetic coupling function and potential\cite{meie_ppn} can be translated into the invariant formalism as\cite{meie2014}
\beq
\label{STG PPN Geff gamma}
G_{\mathrm{eff}} &=& \mathcal{I}_1 \left(1 + \mathcal{I}_5 e^{-m_\Phi r} \right) \,,\qquad
\gamma-1 = -\frac{2e^{-m_\Phi r}}{G_{\mathrm{eff}}}\mathcal{I}_1\mathcal{I}_5 \,,\\
\label{STG PPN beta}
\beta - 1 
&=&\frac{1}{2}\frac{\mathcal{I}_1^3\mathcal{I}_5}{G_{\mathrm{eff}}^2} \frac{\mathcal{I}_5^\prime}{\mathcal{I}_1^\prime}  e^{-2 m_\Phi r} -  \frac{m_\Phi r}{G_{\mathrm{eff}}^2} \mathcal{I}_1^{2}\mathcal{I}_5 \, \beta(r) \,,
\eeq
where $m_\Phi = \frac{1}{\ell}\sqrt{\frac{\mathcal{I}_2^{\prime\prime}}{2\mathcal{I}_1 \left( \mathcal{I}_3^\prime \right)^2}}$ can be understood as an effective mass and $\beta(r)$ is a bit cumbersome radius dependent term.

\section{Scalar Field Fixed Point in FLRW Cosmology without Matter}

The scalar field dynamics in flat Friedmann-Lema\^itre-Robertson-Walker (FLRW) cosmology without matter is given by ($\varepsilon = \pm1$ expanding / contracting universe)\cite{meie2014}
\be
\label{skalaarvalja_kosmo_vorrand_invariantidena}
 \frac{d^2}{d\hat{t}^2}\mathcal{I}_3  = - \varepsilon \sqrt{3\left(\frac{d}{d\hat{t}} \mathcal{I}_3 \right)^2 + \frac{3}{\ell^{2}}\mathcal{I}_2}\,\,\frac{d}{d\hat{t}}\mathcal{I}_3 - \frac{1}{2\ell^2}\frac{d\mathcal{I}_2}{d\mathcal{I}_3} \,,
\ee
where $d\hat{t}=\sqrt{\mathcal{A}} \, dt$.
We may linearize this equation around the fixed point at $\Phi_0$: $\left. \frac{\mathcal{I}_2^\prime}{\mathcal{I}_3^\prime} \right|_{\Phi_0} = 0$. The solutions of the linearized equation are
\be
\label{solution_for_linearized_equation_in_terms_of_invariants}
\mathcal{I}_3(\hat{t}) = M_1 e^{\lambda^{\varepsilon}_{+}\hat{t}} + M_2 e^{\lambda^{\varepsilon}_{-}\hat{t}} \,,
\ee
with the eigenvalues 
$\lambda^{\varepsilon}_\pm = \frac{1}{2\ell}\left[ -\varepsilon \sqrt{3\mathcal{I}_2} \pm \sqrt{3\mathcal{I}_2 - 2 \frac{d^2\mathcal{I}_2}{d\mathcal{I}_3^2}}\right]_{\Phi_0}$. The invariant formalism makes it clear that given a particular STG the existence of the fixed point as well as the physically observable qualities of the approximate solutions (attractor, repeller, etc) and  periods of the oscillations are independent of the parametrization.

To express the solution (\ref{solution_for_linearized_equation_in_terms_of_invariants}) in terms of the scalar field we can Taylor expand 
\be
\Phi(\hat{t}) - \Phi_{0} =
\pm \left. \frac{1}{\mathcal{I}_3'} \right|_{\Phi_0} \mathcal{I}_3(\hat{t})
+
\left.\frac{1}{4}\left(\frac{1}{(\mathcal{I}_3')^2}\right)^\prime\right|_{\Phi_0} \cdot \mathcal{I}_3^2(\hat{t})\,.
\label{taylor_back}
\ee
Note that the fixed point condition can be satisfied in two ways: $\Phi_\bullet$: $\left.\mathcal{I}_2^\prime\right|_{\Phi_\bullet}=0, \left. \frac{1}{\mathcal{I}_3^\prime} \right|_{\Phi_\bullet}\neq 0$, and $\Phi_\star$: $\left. \frac{1}{\mathcal{I}_3^\prime} \right|_{\Phi_\star}=0$.
For example in JF BDBW parametrization the first is equivalent to $\Psi V' - 2 V=0$, while the second to $\frac{1}{\omega} = 0$.
In the first case $\Phi_\bullet$ the solution for $\Phi(\hat{t})$ is linear, but for $\Phi_\star$ the first term in Eq.~(\ref{taylor_back}) vanishes and the solution will be nonlinear.\cite{meie2014} 
However, one must bear in mind that the value of the scalar field itself is not observable, and linear / nonlinear behavior stems from the choice of parametrization. A thorough analysis shows how the solutions for $\Phi_\bullet$ and $\Phi_\star$ actually transform into each other under conformal rescaling and reparametrization.\cite{meie2014talv, meie2015}

\section{Multiscalar-tensor gravity}

Extending the action (\ref{fl_moju}) to $n$ scalar fields $\Phi^A$ gives multiscalar-tensor gravity,\cite{DamourEF,mstg2015}
\begin{align}
\nonumber
S =& \frac{1}{2\kappa^2}\int_{V_4} d^4x\sqrt{-g} \left\lbrace \mathcal{A} \left( \bPhi \right)  R - \mathcal{B}_{AB} \left( \bPhi \right) g^{\mu\nu} \nabla_\mu \Phi^A \nabla_\nu \Phi^B - 2\ell^{-2} \mathcal{V} \left( \bPhi \right)  \right\rbrace \\
\label{MSTG_Flanagan_action}
&+ S_\mathrm{m}\left[ e^{2\alpha \left( \bPhi \right)}g_{\mu\nu}, \chi \right] \,
\end{align}
where we assume $\mathcal{B}(\boldsymbol{\Phi})_{AB}$ to be an invertible symmetric square matrix function of $\boldsymbol{\Phi}=\{ \Phi^A \}$. The action (\ref{MSTG_Flanagan_action}) is invariant under local Weyl rescaling and reparametrization of the scalar fields,
\begin{equation}
\label{conformal_and_scalar_fields_transformations}
g_{\mu\nu} = e^{2\bar{\gamma}\left( \bar{\bPhi} \right)  } \bar{g}_{\mu\nu} \,, \qquad \qquad 
\Phi^A = \bar{f}^A\left( \bar{\bPhi} \right) \,,
\end{equation}
while the arbitrary functions transform as\cite{mstg2015}
\beq
\label{mstg_Flanagan_transformations}
	\mathcal{A}(\bar{\boldf}(\bar{\bPhi})) &=& e^{-2\bar{\gamma}(\bar{\bPhi})}\bar{\mathcal{A}}(\bar{\bPhi}) \,, \nonumber \\
	\mathcal{V}(\bar{\boldf}(\bar{\bPhi})) &=& e^{-4\bar{\gamma}(\bar{\bPhi})} \bar{\mathcal{V}}(\bar{\bPhi}) \,, \nonumber \\
	\alpha(\bar{\boldf}(\bar{\bPhi})) &=& \bar{\alpha}(\bar{\bPhi}) - \bar{\gamma}(\bar{\bPhi}) \,, \nonumber \\
	\nonumber
	\mathcal{B}_{AB}(\bar{\boldf}(\bar{\bPhi})) &=& e^{-2\bar{\gamma}(\bar{\bPhi})} \left( \bar{f}^{C}_{\phantom{C},A} \right)^{-1} \left( \bar{f}^{D}_{\phantom{C},B}  \right)^{-1} \left\lbrace \bar{B}_{CD}(\bar{\bPhi}) - 6 \bar{\gamma}_{,C} \bar{\gamma}_{,D} \bar{\mathcal{A}}(\bar{\bPhi}) \right. \\
	&& \left. + 3 \left( \bar{\gamma}_{,D} \bar{\mathcal{A}}_{,C} + \bar{\gamma}_{,C} \bar{\mathcal{A}}_{,D} \right)  \right\rbrace \,,
\eeq
where the comma $,_A$ denotes partial derivative w.\ r.\ t.\ $\Phi^A$.
Again, by invoking these transformations 
it is possible to express the theory in various parametrizations, e.g.\ the Jordan frame BDBW type or Einstein frame canonical parametrization.\cite{mstg2015}

\section{MSTG Invariants and the Metric for the Space of Scalar Fields}\label{Invariants_and_metric}

Analogously to the single field case, we can construct multiscalar quantities which remain invariant under the  transformations (\ref{conformal_and_scalar_fields_transformations}), like\cite{mstg2015}
\begin{equation}
\label{mstg_invariants}
\mathcal{I}_1(\bPhi) \equiv \frac{e^{2\alpha(\bPhi)}}{\mathcal{A}(\bPhi)} \,, \qquad \qquad \mathcal{I}_2(\bPhi) \equiv \frac{\mathcal{V}(\bPhi)}{ \left( \mathcal{A}(\bPhi) \right)^2 } \,.
\end{equation}
Any function of these will also give an invariant, e.g.\
\begin{equation}
\mathcal{I}_4(\bPhi) \equiv \frac{\mathcal{I}_2}{\mathcal{I}_1^2} = \frac{\mathcal{V(\bPhi)}}{e^{4\alpha(\bPhi)}} \,.
\end{equation}
However, in comparison with the single field case, the third invariant (\ref{I_3}) and the other two rules (b), (c) for constructing further invariants do not directly generalize to the multifield case.

Note that under the scalar reparametrizations alone $\mathcal{B}_{AB}$ transforms as a second order covariant tensor and can be considered to be the metric of the space of fields $\Phi^A$ (like for the $\sigma$-models minimally coupled to gravity). But under the local Weyl rescaling of the spacetime metric it gains additive terms. Therefore it is more reasonable to introduce the metric of the space of fields and its transformation rule as 
\begin{equation}
\label{definition_of_F}
\mathcal{F}_{AB} \equiv \frac{ 2\mathcal{A} \mathcal{B}_{AB} + 3\mathcal{A}_{,A} \mathcal{A}_{,B} }{4\mathcal{A}^2} \,, \qquad  \mathcal{F}_{AB} = \left( \bar{f}^{C}_{\phantom{C},A} \right)^{-1} \left( \bar{f}^{D}_{\phantom{D},B} \right)^{-1} \bar{\mathcal{F}}_{AB} \,.
\end{equation}
We assume $\mathcal{F}_{AB}$ to be an invertible matrix and denote its inverse by $\mathcal{F}^{AB}$. These can be used to formally raise, lower, and contract scalar field indexes, define a covariant derivative in the space of scalar fields, etc.  

Now we can generalize the third invariant $\mathcal{I}_3$ to be an indefinite integral
\begin{equation}
\label{mstg_I_3}
\mathcal{I}_3(\bPhi) \equiv \int \sqrt{\det|\mathcal{F}_{AB}|} \,d\Phi^1\wedge\ldots\wedge d\Phi^n \,.
\end{equation}
In fact, any combination of the fields, which happens to be a ``scalar'' in the space of fields, e.g.\
\begin{equation}
\label{mstg_I_5}
\mathcal{I}_5(\bPhi) \equiv \frac{1}{4} \mathcal{F}^{AB} \left(\ln \mathcal{I}_1 \right)_{,A} \left(\ln \mathcal{I}_1 \right)_{,B}
\end{equation}
remains invariant under the transformations (\ref{conformal_and_scalar_fields_transformations}). 
In this way we may construct many more invariant quantities.\cite{mstg2015}

\section{MSTG PPN Parameters in Terms of Invariants}

By invoking an insightful mapping technique, it is again  possible to reconstruct the expressions for the effective gravitational constant and PPN parameters for MSTG (without potential):\cite{mstg2015}
\beq
	\label{MSTG_PPN_in_generic_frame}
G_{\mathrm{eff}} &=& \frac{\kappa^2}{8\pi} \left[ \mathcal{I}_1  \left( 1 + \mathcal{I}_5 \right) \right]\,, \qquad \qquad
\gamma - 1 = -2 \left( \frac{ \mathcal{I}_5 }{ 1 + \mathcal{I}_5 } \right) \,, \\
\beta - 1 &=& \frac{ \left( \ln \mathcal{I}_1 \right)^{,A} \left( \ln \mathcal{I}_1 \right)^{,B} \left( \left( \ln \mathcal{I}_1 \right)_{,AB} - \frac{1}{2} \mathcal{F}_{AB,C} \left( \ln \mathcal{I}_1 \right)^{,C} \right) }{16 \left( 1 + \mathcal{I}_5 \right)^2} \,.
\eeq
This result reduces to the earlier results in the literature for Einstein frame,\cite{DamourEF} and Jordan frame\cite{Erik2014}, as well as to Eqs.\ (\ref{STG PPN Geff gamma}), (\ref{STG PPN beta}) in the single field case. A further inclusion of the potential is also possible.
Once again we conclude that physical observables are frame and parametrization independent since they transform as invariants.


\section{Outlook}
It would be interesting to apply the formalism of invariants to study the contentious issues of cosmological perturbations and quantum corrections. It would be also interesting to generalize it for STGs and MSTGs with derivative couplings and disformal invariance (Horndeski and beyond).

\section*{Acknowledgments}

This work was supported by the Estonian Science Foundation Grant No. 8837, by the Estonian Ministry for Education and Science Institutional Research Support Project IUT02-27 and by the European Union through the European Regional Development Fund (Project No. 3.2.0101.11-0029).

\end{document}